# The influence of phonon thermal conductivity on thermoelectric figure of merit of bulk nanostructured materials with tunneling contacts

L.P. Bulat,[1] D. Kossakovski,[2] and D.A. Pshenay-Severin[3]

[1]*National Research University ITMO, ul. Lomonosova, 9, St. Petersburg 191002, Russia,*

lbulat@mail.ru

[2]*ZT Plus, 1321 Mountain View Circle, Azusa, CA 91702,*

Dmitri.Kossakovski@ztplus.com

[3]*Physicotechnical Institute, Russian Academy of Sciences, ul. Politekhnicheskaya 26, St. Petersburg, 194021, Russia,* d.pshenay@mail.ru

A composite material is considered which consists of conducting nanoparticles separated by tunneling dielectric barriers. The influence of the phonon thermal conductivity of dielectric matrix $\kappa_d$ on the thermoelectric figure of merit of this composite material is theoretically investigated. The range of $\kappa_d$ values and barrier parameters that can lead to the thermoelectric figure of merit greater than unity is estimated. The influence of space charge and nonlinearity of current-voltage relations of tunneling barrier are also discussed.

## I. Introduction

Thermoelectric materials are used in heat to electricity converters and refrigerators. Their main advantages are the lack of moving parts and environmentally unsafe refrigerants, maintenance-free operation and usage in waste heat utilization.[1-2] The efficiency of energy conversion is determined by the thermoelectric figure of merit $Z$

$$ZT = \frac{S^2 \sigma}{\kappa}, \qquad (1)$$

where $T$ is the absolute temperature, $S$ is the Seebeck coefficient, $\sigma$ and $\kappa$ are electrical and thermal conductivities of the material. One of the possibilities to enhance ZT above unity proposed recently is to use bulk nanostructured materials.[3-7] These composites are synthesized by means of ball milling of initial BiTe-SbTe solid solution material and subsequent hot-pressing.[3] The final samples were polycrystalline with grains of the size of about 20-30nm[3]. The other authors obtained solid solutions based on PbTe-SbTe with nano-inclusions of Ag or Pb.[6,7] One can imagine that during the preparation process the nanograins can be coated with a dielectric material. So the composite will consist of conducting grains separated by a dielectric matrix. This type of nanostructured material is considered in the present paper.

The conduction in such structures is determined by electron tunneling through the dielectric barriers. This mechanism is similar to that in thermionic energy converters that were considered for the cases of layered geometry[8-10] and for the case of vacuum barrier between conical tip and semiconducting plate.[11] In the previous work[12] the influence of three dimensional geometry of the grains on the thermoelectric efficiency of such nanocomposite was considered for the case of vacuum barriers. The grain geometry was modeled by two truncated cones with the same base (fig. 1). It was shown that in the case of vacuum barriers the thermoelectric efficiency of composite material can reach the values $ZT=3.0-4.0$ at room temperature.[12] In this work the results of the previous consideration[12] are extended to take into account the lattice thermal conductivity of intergrain dielectric media $\kappa_d$. The range of $\kappa_d$ values and parameters of tunneling barrier that gave $ZT > 1$ were examined. In addition the influence of the space charge inside the barrier and possibility of nonlinearity of current-voltage dependence of tunneling junction are discussed.

## II. Calculation of electric and heat currents in tunneling junction

In the calculation it was assumed that the tunneling current flows only through the cylindrical part with the height $d$ and base radius $r_0$ and the energy height of barrier is $\varepsilon_b$. The total energy $\varepsilon$ of electron is conserved during the trip across the gap. Due to the conservation of the component of the momentum parallel to the electrode surface $k_\parallel$ the component of the energy corresponding to the movement perpendicular to that surface $\varepsilon_x$ is conserved as well. This value is a sum of kinetic energy and potential energy inside the barrier

$$\varepsilon_x = \frac{\hbar^2 k_x^2(x)}{2m} + U(x). \tag{2}$$

The energy is considered to be zero at the bottom of the conduction band of the first electrode.

The number of electrons with a given energy $\varepsilon_x$ inside the interval $dk_x$ can be calculated as

$$\frac{dN(x)}{dk_x} = \frac{2}{V}\sum_{k_\parallel} f_0(\varepsilon_\parallel + \varepsilon_x - \mu) = \frac{mk_0 T}{2\pi^2 \hbar^2}\ln\left(1 + e^{-(\varepsilon_x^* - \mu^*)}\right) \tag{3}$$

Here, $\mu$ is the chemical potential, $m$ is the electron effective mass, $f_0$ is the Fermi-Dirac distribution function and all variables with asterisks are measured in $k_0 T$ units. The same quantity per unit interval of energy is

$$\frac{dN_x(x)}{d\varepsilon_x^*} = \frac{(mk_0 T)^{3/2}}{2^{3/2}\pi^2 \hbar^3 \sqrt{\varepsilon_x^* - U^*(x)}} \ln\left(1 + e^{-(\varepsilon_x^* - \mu^*)}\right). \tag{4}$$

The contribution of the electrons in the energy interval $(\varepsilon_x, \varepsilon_x + d\varepsilon_x)$ to the current density flow equals to

$$d j_x = -e v_x \, d N_x(x) = -e \frac{m(k_0 T)^2}{2\pi^2 \hbar^3} \ln\left(1 + e^{-(\varepsilon_x^* - \mu^*)}\right) d\varepsilon_x^*. \qquad (5)$$

The current flow in the vertical direction (x-axis) inside the tunneling gap is given by the following equation

$$j_x = -e \frac{m(k_0 T_1)^2}{2\pi^2 \hbar^3} \int_0^\infty D(\varepsilon_x^*) \nu(\varepsilon_x^*, \mu_1^*) \left(1 - \frac{T_2}{T_1} \frac{\nu(\varepsilon_x^*, \mu_2^*)}{\nu(\varepsilon_x^*, \mu_1^*)}\right) d\varepsilon_x^*, \qquad (6)$$

where indices $i = 1, 2$ denote two electrodes, $D(\varepsilon_x^*)$ is the electron tunneling probability and

$$\nu(x, y) = \ln\left(1 + e^{-(x-y)}\right). \qquad (7)$$

The equation for the heat flow is similar to (6)

$$q_x = \frac{m(k_0 T_1)^3}{2\pi^2 \hbar^3} \int_0^\infty D(\varepsilon_x^*) \theta(\varepsilon_x^*, \mu_1^*) \left(1 - \left(\frac{T_2}{T_1}\right)^2 \frac{\theta(\varepsilon_x^*, \mu_2^*)}{\theta(\varepsilon_x^*, \mu_1^*)}\right) d\varepsilon_x^*, \qquad (8)$$

where

$$\theta(x, y) = \frac{1}{6}\left(\frac{\pi^2}{2} + 3(x-y)^2 + 6x \ln\left(1 + e^{y-x}\right)\right) + \mathrm{Li}_2\left(-e^{x-y}\right), \qquad (9)$$

and $\mathrm{Li}_2(x)$ - dilogarithm.

In an external electric field initial square barrier changes to triangular. For the calculation of current and heat flow we use tunneling probability in WKB approximation for triangular barrier:[13]

$$D_{tri}(\varepsilon_x) = \begin{cases} \exp\left(-\frac{4}{3}\sqrt{\frac{2m}{\hbar^2}} \frac{(\varepsilon_b - \varepsilon_x)^{3/2} - (\varepsilon_b - Fd - \varepsilon_x)^{3/2}}{F}\right), & \varepsilon_x < \varepsilon_b - Fd \\ \exp\left(-\frac{4}{3}\sqrt{\frac{2m}{\hbar^2}} \frac{(\varepsilon_b - \varepsilon_x)^{3/2}}{F}\right), & \varepsilon_b - Fd < \varepsilon_x < \varepsilon_b \\ 1, & \varepsilon_x > \varepsilon_b \end{cases} \qquad (10)$$

In this equation $F = -eE$ is the force acting on the electron in electrical field $E$.

When temperature difference is smaller than the average temperature $|\Delta T| = |T_2 - T_1| \ll \overline{T}$ and the potential drop on the single barrier $\mu_2 - \mu_1 = -e\Delta V$ is small $|e\Delta V| \ll k_0 \overline{T}, \varepsilon_b$, linear transport coefficients can be obtained neglecting the change of the barrier shape. For example, electric current density can be written as [14, 15]

$$j_x = \int_{v_x>0} \frac{2 d^3 \mathbf{k}}{(2\pi)^3} \left\{ -e v_x(\mathbf{k}) \, D(\mathbf{k}) \left( -\frac{\partial f_0(\varepsilon^* - \mu_1^*)}{\partial \varepsilon^*} \right) \left( \frac{e\Delta V}{k_0 T_1} - (\varepsilon^* - \mu_1^*) \frac{\Delta T}{T_1} \right) \right\}. \quad (11)$$

The equation for heat current density $q_x$ can be obtained from (11) replacing $-e v_x(\mathbf{k})$ with $(\varepsilon - \mu_1) v_x(\mathbf{k})$. It is useful to introduce an integral

$$J_n = \int_{v_x>0} \frac{2 d^3 \mathbf{k}}{(2\pi)^3} \left\{ (\varepsilon^* - \mu_1^*)^n \, v_x(\mathbf{k}) \, D(\mathbf{k}) \left( -\frac{\partial f_0(\varepsilon^* - \mu_1^*)}{\partial \varepsilon^*} \right) \right\}. \quad (12)$$

Then expressions for tunneling electrical conductivity $\sigma_t$, thermopower $S_t$ and thermal conductivity at zero voltage drop $\kappa_{t,\Delta V=0}$ can be written as

$$\sigma_t = (e^2 / k_0 T_1) J_0, \quad S_t = (e/T_1) J_1 / \sigma_t, \quad \kappa_{t,\Delta V=0} = k_0 J_2. \quad (13)$$

Usual thermal conductivity at zero electric current can be expressed as $\kappa_t = \kappa_{t,\Delta V=0} - S_t^2 \sigma_t T_1$ [15].

For the case considered here, when tunneling probability depends only on $\varepsilon_x$ and effective masses in electrodes and barrier are the same integration in (12)-(13) can be partially performed analytically. These equations were obtained in the previous work[12] and are given in appendix.

Let's note that the units of barrier kinetic coefficients $\sigma_t$ and $\kappa_t$ are different from that of a bulk sample as they connect flow density with potential and temperature difference and not with potential and temperature gradients. For comparison with the bulk values it is convenient to use values of $\sigma_t d$ and $\kappa_t d$.

The dependence of current density on voltage drop is shown on the fig. 2. In the calculations the effective mass was assumed to be equal to that of free electron $m_0$. The figure shows that there is a wide range of voltages for which junction operates in linear region. Usually, the practical current densities[8] start from $1 A/cm^2$. On the figure 2 it can be seen that such currents are attainable for the barrier height of several tenths of eV and barrier thickness of several nanometres even in linear regime of operation. Though the smallest available work function up to date[16-18] is about 0.8eV, in heterostructures the barrier height $\varepsilon_b$ is determined by the difference in the energy positions of conduction (valence) band in material of grains and matrix which can be smaller (of the order of 0.1eV). So, the values of $\varepsilon_b \sim 0.1 eV$ look quite reasonable.

On the figure 3 the plot of Peltier coefficient $\Pi_t = q/j$ is presented. It can be seen that the values of Peltier coefficient are greater than conventional values for room temperature thermoelectrics based on $Bi_2Te_3$-$Sb_2Te_3$ solid solutions. The values of Peltier coefficient increase when the relative contribution of more energetic carriers to the heat flow increases. The energy filtering of carriers in the linear region of operation is determined by the exponential dependence of tunneling probability on carrier energy. So, the contribution to the heat flow from carriers with energies lower than $\varepsilon_b$ is small. At larger voltages the shape of barrier changes from square to triangular. Generally, this can lead to stronger energy filtering and increase in Peltier coefficient. But this effect can be seen only for large $\varepsilon_b$ (see 5 on fig. 3). For smaller $\varepsilon_b$ this effect is less important and the increase of tunneling probability for all electrons with energies lower than $\varepsilon_b$ leads to decrease in Peltier coefficient (see curves 1-4 on fig. 3).

In a bulk composite material with grains of 20-30nm in size the voltage drop and temperature difference are small, as was discussed in the previous work[12]. For example, if the size of the sample is 1mm, and the grain size is 20nm, then $\Delta T$ is $5 \cdot 10^4$ times smaller than total temperature difference of the order of 100K, so $\Delta T \sim 2 \cdot 10^{-3} K << \overline{T}$. This can lead to the total voltage difference due to Seebeck effect of the order of 0.1V for large $S = 10^3 \mu V/K$ and $\Delta V \sim 2\mu V$. So, even for the several orders of magnitude increase in total voltage difference the single junction will operate in linear regime. As can be seen from fig. 2-3 linear region of operation can be attractive for getting large Peltier coefficients and reasonable current densities. Therefore, only this region will be considered.

The dependences of the electrical conductivity and thermopower on the barrier thickness in the linear operation region are plotted on the figure 4, as calculated using equations from the previous work[12]. It can be seen that for small $d$ the electrical conductivity decreases with the increase of $d$ due to decrease of the tunneling probability and then it slowly increases as it is proportional to barrier thickness for ballistic transport. The thermopower, on the contrary, increases with $d$ due to better energy filtering of heat carriers.

The potential barrier can be changed not only due to external electric field but also due to charge buildup and image charge effects. The latter effect decreases the energy height of the potential barrier that leads to increase in current density. However, the former effect leads to the increase of the barrier height. The influence of both effects is less pronounced in dielectric media with permittivity $\varepsilon_d > 1$. In order not to complicate the treatment they are not included in consideration. We just estimate the possible influence of the charge buildup effect as the one that worsens the situation and show that for small $d$ this effect is negligible. The effect is connected with the electrons which have enough energy to penetrate into barrier region and create negative charge that prevents other electrons from getting there. This leads to effective increase of the barrier height and decrease of the current density. This effect was taken it into account in [8, 19] for the case of thermionic emission and classical statistics. The simple equation for barrier form in equilibrium given in [8] reads

$$U(x) = \varepsilon_b + 2k_0 T \ln[\cos((x-d/2)c/2x_l)/c], \qquad (14)$$

where $x_l^2 = \left(\varepsilon_d \sqrt{\pi} \hbar^3 / 4e^2 \sqrt{2k_0 T} m^{3/2}\right) e^{\varepsilon_b^* - \mu^*}$ and constant $c$ can be determined from equation $\cos(dc/4x_l) = c$ for $0 < c < 2\pi x_l / d$. This effect is most important for small $\varepsilon_b$. But it appears that for small barrier thickness the change of the barrier height is less than several percent. For example for $\varepsilon_b = 0.1 \text{eV}$ and $d = 2,5,10 \text{nm}$ the increase in barrier height is 0.23, 1.4 and 5 percent correspondingly. In these estimations $\varepsilon_d$ is taken to be unity. In dielectric media $\varepsilon_d > 1$ so the barrier increase will be even less. So, for linear operation region this effect can be disregarded.

**III. Kinetic coefficients of composite medium**

The effective coefficients for composite material are calculated for the model similar to the previous one.[12] The nanoparticle is modeled with the help of two truncated cones with the same base. The nanocomposite is formed from the primitive cells depicted on the figure 1. The geometric parameters are the following: $a$ is the height of each truncated cone, so $2a$ is the size of nanoparticles along the vertical direction. Radii $r_0$ and $r_1$ are the smaller and larger radii of the cone bases and $2\theta$ is the cone aperture angle. The primary cell has a square base in horizontal plane with size $2b$. The height of the elementary cell is $h = 2a + d$. The analyti-cal solution for effective kinetic coefficients was obtained previously but the lattice thermal conductivity of dielectric $\kappa_d$ was not taken into account.[12]

Due to the complexity of the geometry of the considered object and the fact that $\kappa_d \neq 0$ the variables cannot be separated and the analytical solution for kinetic coefficients was not found. In the present work the numerical solution of equations for heat and current flow was used for calculation of the effective kinetic coefficients. In the following treatment the index $n$ for kinetic coefficient $\sigma_n, S_n, \kappa_n$ corresponds to nanograin.

In our calculations the system of differential equations for the temperature and electrical potential distribution was solved numerically:[20]

$$\begin{aligned} &\text{div}(-\sigma\alpha\nabla T) + \text{div}(-\sigma\nabla\phi) = 0, \\ &\text{div}(-(\sigma\alpha^2 T + \kappa)\nabla T) + \text{div}(-\sigma\alpha T\nabla\phi) = \sigma((\nabla\phi)^2 + \alpha\nabla T\nabla\phi), \end{aligned} \quad (15)$$

where $\phi$ is electrical potential.

To calculate the effective coefficients of the nanostructures material one has to set boundary conditions. The set of 5 elementary cells stacked in vertical direction and connected to metallic contacts were considered and in horizontal direction the structure was considered to be periodic. The comparison with the case of only one cell showed that the influence of contacts is negligible. The boundary conditions on outer boundaries are the following: on the left, right, front and back sides of the primary cell normal components of electrical and heat currents are set to zero $j_n = 0, q_n = 0$; on the top $T_1 = 300\,\text{K}, \phi_1 = 0$; on the bottom side one can fix either temperature and potential $T_0, \phi_0$ or inflowing heat and current fluxes $j_n = j_0, q_n = q_0$. Here, the second type of conditions was used because they provide better convergence during numerical calculations. The boundary conditions on the inner boundaries are the continuity of the normal components of current and heat fluxes determined and continuity of temperature and potential fields. These boundary conditions automatically include the Peltier effect at the contact of two dissimilar materials.

For each set of parameters two calculations were performed. In the first run the heat flow density was fixed $q_0$ and $j_0 = 0$. Then the temperature $T_0$ and potential $\phi_0$ on the bottom contact were calculated. From this run the effective thermal conductivity and thermopower can be calculated $\kappa_{eff} = -q_0/(T_1 - T_0)L$, $\alpha_{eff} = -(\phi_1 - \phi_0)/(T_1 - T_0)$. Here $L$ is the size of the considered part of the sample along the vertical direction. Next we perform another calculation with $q_0 = 0$ and fixed $j_0$ and determine the electrical conductivity $\sigma_{eff} = -j_0/(\phi_1 - \phi_0 + \alpha_{eff}(T_1 - T_0))L$.

In the calculations two possible material sets were considered. In the first one the grains were assumed to consist of typical thermoelectric semiconducting material with $\mu = 0$, $\sigma_n = 1000\,\text{Sm/cm}$, $\alpha_n = 200\,\mu\text{V/K}$, $\kappa_{n,ph} = 1\,\text{W/mK}$. These values are close to transport coefficients of p-$Bi_2Te_3$ in cleavage plane (normal to trigonal axis). In the second parameter set the nanograin was assumed to consist of metal (Ag) with $\mu = 5.49\,\text{eV}$, $\sigma_n = 6.3 \cdot 10^5\,\text{Sm/cm}$, $\alpha_n = 1.33\,\mu\text{V/K}$, $\kappa_n = 430\,\text{W/mK}$. The effective coefficients were calculated for the following geometries: $2a = 10, 20, 30\,\text{nm}$, $d = 1, 2, 5\,\text{nm}$, $\varepsilon_b - \mu = 0.05, 0.1, 0.2\,\text{eV}$, $b = 1.1a$, $\theta = 15°, 30°$. Note that for metal nanoparticles $\varepsilon_b$ is counted from the chemical potential level.

Typical dependencies of effective thermoelectric figure of merit on dielectric thermal conductivity are presented in figure 5 for $\theta = 15°$ and $2a = 20\,\text{nm}$. The dashed curves correspond to metallic grains and solid curves correspond to semiconducting ones. From the figure 5 it can be seen that the thermoelectric efficiency can be greater than unity if thermal conductivity of dielectric $\kappa_d < 0.01 \div 0.02\,\text{W/mK}$ at $\varepsilon_b = 0.1\,\text{eV}$ and for $\kappa_d < 0.05\,\text{W/mK}$ at bar-rier height of 0.05eV. The increase in electrical conductivity is preferable for increase of $Z_{eff}T$. Hence $Z_{eff}T$ is greater for smaller $\varepsilon_b$. It is interesting that for sufficiently small $\varepsilon_b$ the electrical conductivity $(\sigma_t d)$ increases with the increase of barrier thickness as it should be for ballistic transport. This leads to larger $Z_{eff}T$ values for larger $d$ (compare curves 1 and 2 on fig. 5). Further increase of $Z_{eff}T$ with $d$ is limited by the number of factors, e.g. due to increase of space charge effect or transition from ballistic to diffusive transport in barriers where the approach developed in this paper is inapplicable.

The comparison of $Z_{eff}T$ for semiconducting and metallic grains shows their similarity. As the main contribution to the thermal conductivity is due to tunneling barrier, the effective Seebeck coefficient is not very sensitive to the value of $S_n$. But the thermal conductivity of nanograin should be increased so that the temperature difference on the tunneling junction is increased as well. This leads to larger $Z_{eff}T$ values for metallic nanoinclusions (compare solid and dashed curves on fig. 5).

It is interesting also to estimate the influence of the anisotropy of semiconducting material on the obtained results. The anisotropy can enter the calculations in two ways: due to anisotropy of band structure and the anisotropy of transport coefficients. $Bi_2Te_3$ is layered material of rhombohedral symmetry. Usually trigonal axis normal to cleavage planes is denoted as crystallographic direction 3. If axis 1 lies in the plane and directed along one of binary axes, then the remaining 2nd direction lies in one of mirror planes. The anisotropy of transport coefficients for holes in $Bi_2Te_3$ are $\sigma_{11}/\sigma_{33} = 2.7$ and $\kappa_{ph,11}/\kappa_{ph,33} = 3$, while thermopower is isotropic [21]. The band structure of $Bi_2Te_3$ can be described by 6-ellipsoidal Drabble-Wolfe model[22]. The effective masses of holes $m_1 = 0.73m_0$, $m_2 = 0.064m_0$, $m_3 = 0.196m_0$ and tilt angle $\theta = 39.6°$ are given in [23]. The probability of tunneling through the square potential barrier for the case of anisotropic energy spectrum is given in [24]. Let's assume that the anisotropic energy spectrum for one ellipsoid can be written as $\varepsilon^{(n)} = (\hbar^2/2m_0)\mathbf{k}^{(n)} \cdot \boldsymbol{\alpha}^{(n)} \cdot \mathbf{k}^{(n)}$, where $\boldsymbol{\alpha}^{(n)}$ is inverse effective mass tensor in crystal axes and indices $n = 1,2$ are for semiconductor and barrier respectively. Components of the wave vector parallel to interface $k_{2(3)}$ and total energy $\varepsilon$ conserve during tunneling. Tunneling probability can be written as [24]

$$D(\mathbf{k}) = \begin{cases} \left(1 + \dfrac{(\xi_1^2\, \alpha_{11}^{(1)}/\alpha_{11}^{(2)} + \xi_2^2\, \alpha_{11}^{(2)}/\alpha_{11}^{(1)})^2}{4\xi_1^2 \xi_2^2} \operatorname{sh}^2(\xi_2 d)\right)^{-1}, & \xi_2^2 > 0 \\ \left(1 + \dfrac{(\xi_1^2\, \alpha_{11}^{(1)}/\alpha_{11}^{(2)} - |\xi_2^2|\alpha_{11}^{(2)}/\alpha_{11}^{(1)})^2}{4\xi_1^2 |\xi_2^2|} \sin^2(|\xi_2| d)\right)^{-1}, & \xi_2^2 < 0 \end{cases} \quad (18)$$

Here the following notation was used $\xi_1^2 = \left(2m_0/\hbar^2\, \alpha_{11}^{(1)}\right)\varepsilon_x^{(1)}$, $\xi_2^2 = \left(2m_0/\hbar^2\, \alpha_{11}^{(2)}\right)\left(\varepsilon_b - \varepsilon_x^{(2)}\right)$ and $\varepsilon_x^{(n)} = \varepsilon - \left(\hbar^2/2m_0\right) \sum_{i,j=2,3} \left(\alpha_{i,j}^{(n)} - \alpha_{i,1}^{(n)}\alpha_{1,j}^{(n)}/\alpha_{11}^{(n)}\right)k_i k_j$.

The difficulty in calculations of tunneling transport coefficients for this case comes from the fact that tunneling probability $D(\mathbf{k})$ depends not only on energy $\varepsilon_x$ in the direction of tunneling but on all components of wave vector $\mathbf{k}$ and the expressions for transport coefficients (12)-(13) involve triple integration. To estimate the influence of anisotropy we calculated tunneling transport coefficients for three possible orientations when $x$ axis of tunneling was directed along one of three crystallographic directions mentioned above. The effective mass in barrier was assumed to be $m_0$ for comparison with previous estimations. Tunneling transport coefficients for these cases are listed in table for barrier parameters $d = 5$ nm and $\varepsilon_b = 0.1$ eV.

|  | $S_{t,ii}$, μV/K | $\sigma_{t,ii} d$, Sm/cm | $\kappa_{t,ii} d$, W/mK |
|---|---|---|---|
| isotropic WKB approximation | 504 | 4.47 | 0.002 |
| $i=1$ | 518 | 6.94 | 0.0029 |
| $i=2$ | 516 | 6.76 | 0.0028 |
| $i=3$ | 504 | 7.43 | 0.003 |

From the table it can be seen that the thermopower in all tree cases differs insufficiently from WKB approximation. When we took into account the influence of anisotropy of effective mass in the semiconductor and its difference from barrier effective mass, tunneling current from single ellipsoid became smaller than in isotropic case. But for considered material when *x* axis is parallel to directions 1 or 2 there are two sets of 2 and 4 equivalent ellipsoids, while when *x* is parallel to trigonal axis all 6 ellipsoids are equivalent. As a result total electrical and thermal conductivities appeared to be 1.5-1.7 times larger than in isotropic WKB case. Tunneling transport coefficients are almost isotropic and for directions 1 and 2 are almost equal.

The influence of the anisotropy of transport coefficients of $Bi_2Te_3$ is illustrated on the figure 5. Curves 2', 2" are plotted for directions 1 and 3 mentioned above taking into account anisotropy of both semiconductor and tunneling transport coefficients. These curves can be compared to curve 2 obtained in isotropic WKB approximation. From the figure it can be seen that the anisotropy of semiconducting material did not change qualitatively the results of estimations. The main impact on the results of estimations comes from the change of tunneling probability and the anisotropy of semiconductor transport coefficients is less important.

In the figure 6 the dependence of effective thermoelectric figure of merit on the geometric parameters is illustrated. For getting larger $Z_{eff}T$ values the smaller grain size is preferable to increase the barrier contribution to Seebeck coefficient (compare 1 and 2 on fig. 6). In the figure 6 the curve 4 is plotted for layered geometry and it can be seen that in this case larger $Z_{eff}T$ values can be obtained for the same $\kappa_d$. This is because in the layered geometry the area of tunneling contact increases which leads to the increase of effective electrical conductivity. The decrease of the thermoelectric figure of merit for the larger $\theta$ with the other parameters unchanged is due to the same reason (curve 3 on fig.6).

## IV. Conclusions

In the present work effective transport coefficients and thermoelectric figure of merit was calculated for nanostructured composite material that consists of conductive grains separated by a dielectric matrix. The tunneling current and heat flow through the dielectric barrier was calculated for linear and nonlinear regions of operation. It was shown that the current density can be larger than $1A/cm^2$ even in the linear operating region if barrier height $\varepsilon_b$ is less than about 0.2eV and barrier width $d$ is less than about 5nm. Estimations showed that at such small barrier widths the increase of the barrier height due to the free electron space charge inside the barrier is negligible.

A range of the lattice thermal conductivity of dielectric that can lead to $Z_{eff}T > 1$ was estimated. This range depends on the parameters of tunneling barriers and grain size, e.g. for barriers with the thickness of 5nm and height 0.05-0.1eV $\kappa_d$ values should be less than 0.02-0.05 W/m K. This result was compared to the layered geometry and it was shown that for the same barrier parameters the upper limit for $\kappa_d$ is expanded to 0.1 W/m K. Thus there are three types of requirements that should be satisfied in order to get $Z_{eff}T > 1$. The first one is the thickness of barrier to be less than 5nm – this can be readily satisfied with current level of material technology. The second requirement is the low thermal conductivity of barrier (less than 0.05-0.1W/m K). The third one is that the barrier height is less than 0.1eV.

The second and third conditions can at the first glance seem to be incompatible. To resolve this contradiction one of the possibilities is to use porous materials with low thermal conductivity. For example in sintered aerogel films (porous $SiO_2$) the thermal conductivity can be well below 0.1 W/m K.[25] But in the case of porous material the tunneling barrier height is determined by the work function that could hardly be less than 0.8eV.[16-18] In this case if the pores are filled with a gas then even tiny thermal conductivity of gaseous phase would completely cancel the effect of thermoelectric efficiency increase due to tunneling through the barrier structure. The contradiction of the second and the third requirement for this case can be resolved only with vacuum barriers. The vacuum is a unique dielectric material that can satisfy all requirements. So, one of the possibilities is that corresponding thermoelectric material with good efficiency could be a porous nanostructure based on bismuth telluride with vacuum pores.

The second possibility that has emerged recently is to use dense materials with ultralow thermal conductivity.[26] Usually the low limit of thermal conductivity is connected with disordered materials[27] where the mean free path for phonons is about interatomic distances. But in the multilayered $WSe_2$ material thermal conductivity as low as 0.05 W/m K was observed.[26] It is interesting that in this material low thermal conductivity is connected with the phonon scattering on the precisely ordered Se-W-Se layers connected with weak van der Waals forces. Taking into account that bismuth telluride also consists of layers connected by van der Waals forces it possible to imagine the preparation of at least layered structure from thin barrier $WSe_2$ layers with low thermal conductivity and layers of bismuth telluride as a goods thermoelectric material. For the bulk material electron affinity and band gap of $WSe_2$ are about 4eV and 1.2eV correspondingly.[28] For $Bi_2Te_3$ the electron affinity is similar about 4.125-4.525eV.[29] This suggests that if the proper doping levels are achievable then one can obtain the work function difference and corresponding barrier height about 0.1-0.05eV. To confirm this possibility further investigation of the band diagram and transport properties of this type of structure are required.

**Acknowledgement**

The work was supported by the ZT Plus-Amerigon.

**Appendix**

The expressions for tunneling transport coefficients obtained in [12] can be written as follows:

$$\sigma_t = \frac{e^2 \, m \, k_0 \, T}{2 \pi^2 \hbar^3} \int_0^\infty D(\varepsilon_x^*) f_0(\xi) d\varepsilon_x^* , \tag{A1}$$

$$|\beta_t| = \frac{e \, m \, k_0^2 \, T}{2 \pi^2 \hbar^3} \int_0^\infty D(\varepsilon_x^*) \left(\xi f_0(\xi) + \ln\left(1 + e^{-\xi}\right)\right) d\varepsilon_x^* , \tag{A2}$$

$$\kappa_{t,\Delta V=0} = \frac{m \, k_0^3 \, T_1^2}{2 \pi^2 \hbar^3} \int_0^\infty D(\varepsilon_x^*) \left(\xi^2 f_0(\xi) + 2\xi F_0(-\xi) + 2 F_1(-\xi)\right) d\varepsilon_x^* . \tag{A3}$$

In these expressions $\xi$ is defined as $\xi = \varepsilon_x^* - \mu_1^*$. Fermi integrals are defined as $F_n(y) = \int_0^\infty f_0(x-y) x^n dx$. It can be shown that $F_n(y) = -\Gamma(n+1) \text{Li}_{n+1}(-e^y)$, where $\Gamma(n+1)$ and $\text{Li}_n(x)$ are gamma-function and poly-logarithm. Using this relation the expression for $\kappa_{t,\Delta V=0}$ was rewritten in more compact form compared to [12].

[18] G.G. Magera, P.R. Davis. J. Vac. Sci. Technol. **A 11**, 2336 (1993).

[19] J.B. Scott. J. Appl. Phys. **52**, 4406 (1981)

[20] M. Jaegle. "Multiphysics Simulation of Thermoelectric Systems - Modeling of Peltier-Cooling and Thermoelectric Generation" in Proceedings of the COMSOL Conference 2008 Hannover, http://www.comsol.com/papers/5256

[21] B. M. Goltsman, V. A. Kudinov, and I. A. Smirnov, *Thermoelectric Semiconductor Materials Based on $Bi_2Te_3$* (Nauka, Moscow, 1972; Army Foreign Science and Technology Center, Charlottesville, Virginia, United States, 1973)

[22] J. R. Drabble and R. Wolfe, Proc. Phys. Soc., London, Sect. B 69, 1101 (1956).

[23] M. Stordeur, M. Stoelzer, H. Sobotta, and V. Riede, Phys. Status Solidi B 150, 165 (1988).

[24] K.-Y. Kim, B. Lee, Phys. Rev. **B 58**, 6728 (1998)

[25] A. Jain, S. Rogojevic, Sh. Ponoth, W.N. Gill, J.L. Plawsky, E. Simonyi, Sh.-T. Chen, P.S. Ho. J. Appl. Phys. **91**, 3275 (2002).

[26] C. Chiritescu, D.G. Cahill, N. Nguyen, D. Johnson, A. Bodapati, P. Keblinski, P. Zschack. Science **315**, 351 (2007).

[27] D. G. Cahill, S. K. Watson, R. O. Pohl, Phys. Rev. **B 46**, 6131 (1992).

[28] O. Lang, Y. Tomm, R. Schlaf, C. Pettenkofer, W. Jaegermann. J. Appl. Phys. **75**, 7814 (1994).

[29] J. Nagao, E. Hatta, K. Mukasa. Proceedings of the XV International Conference on Thermoelectrics. Pasadena, CA, USA, 1996, p. 404.
18

# Figures

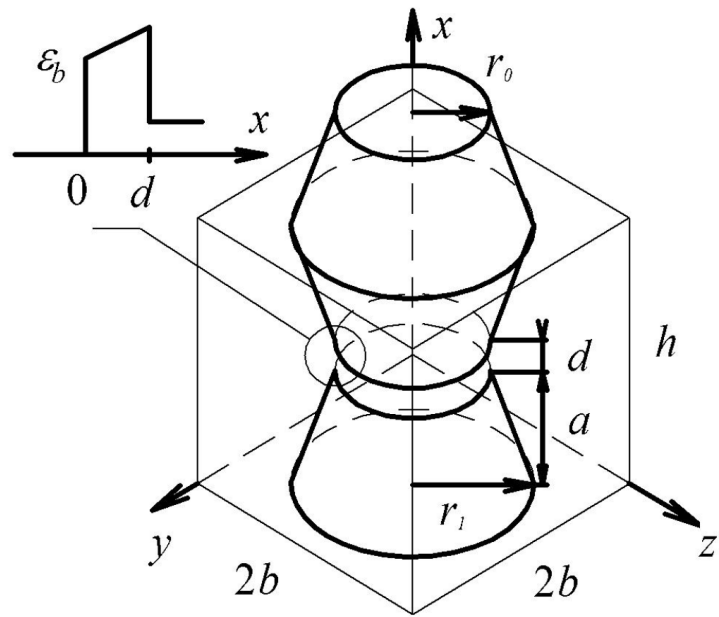

Fig. 1. Schematic drawing of the nanoparticle modeled as two truncated cones. $2a$ - nanoparticle length, $r_0$ - the radius of truncated part, $r_1$ - the radius of base of the cone, $h$ and $2b$ - height and width of elementary cell, $d$ - width of tunneling barrier. In the inset the tunneling barrier of height $\varepsilon_b$ under applied voltage difference is schematically drawn.

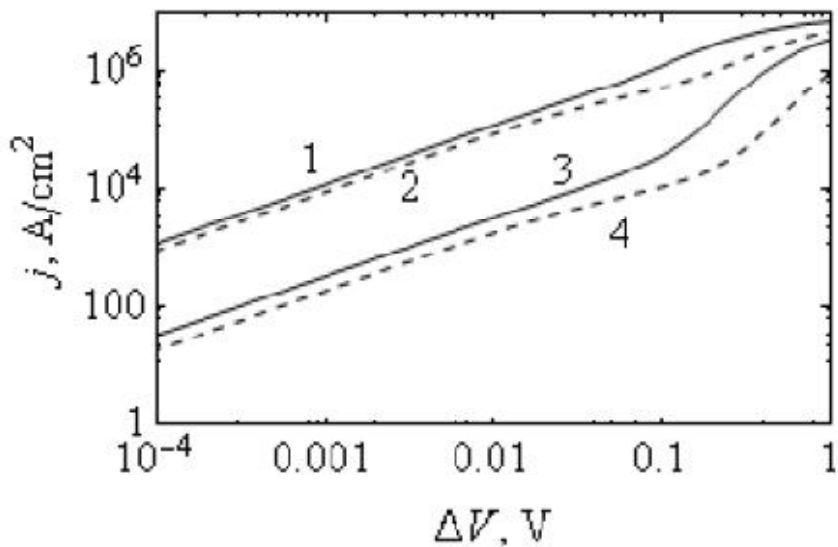

Fig. 2. Current-voltage dependence of tunneling contact. $\varepsilon_b = 0.1\,\text{eV}$ (1,2), $0.2\,\text{eV}$ (3,4); $d = 2\,\text{nm}$ (1,3), $5\,\text{nm}$ (2,4).

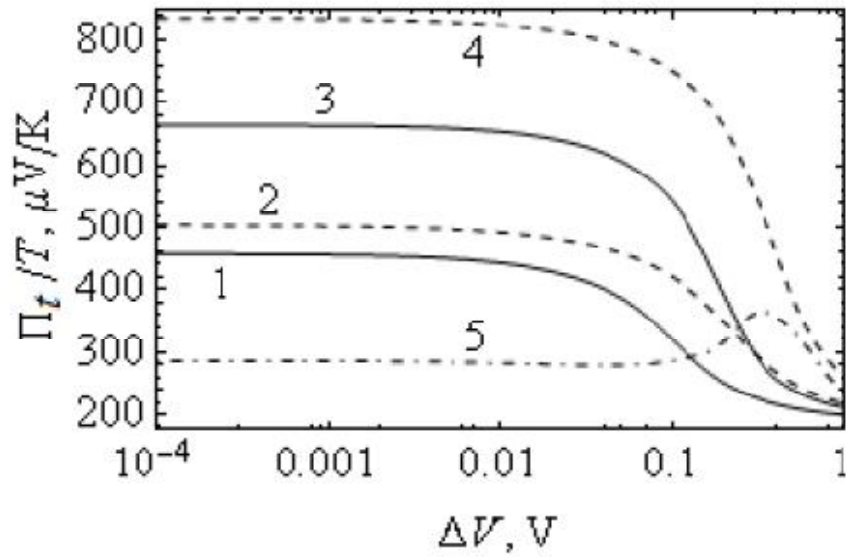

Fig. 3. The dependence of Peltier coefficient of tunneling contact on voltage. (1-4 - see fig. 2; 5 - $\varepsilon_b = 0.5\,\text{eV}$, $d = 2\,\text{nm}$).

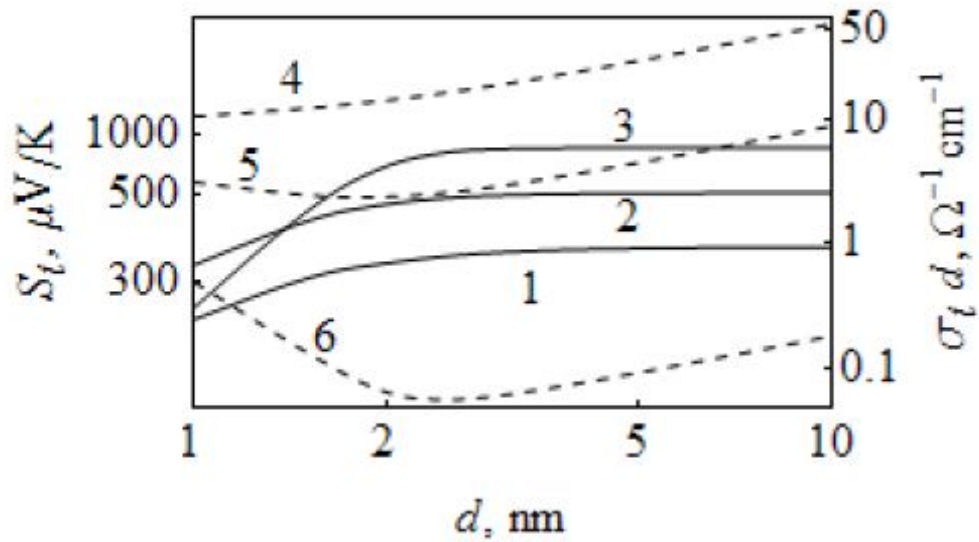

Fig. 4. The dependence of Seebeck coefficient (1-3) and electrical conductivity (4-6) on barrier thickness. $\varepsilon_b$=0.05eV (1,4); 0.1eV (2,5); 0.2eV (3,6).

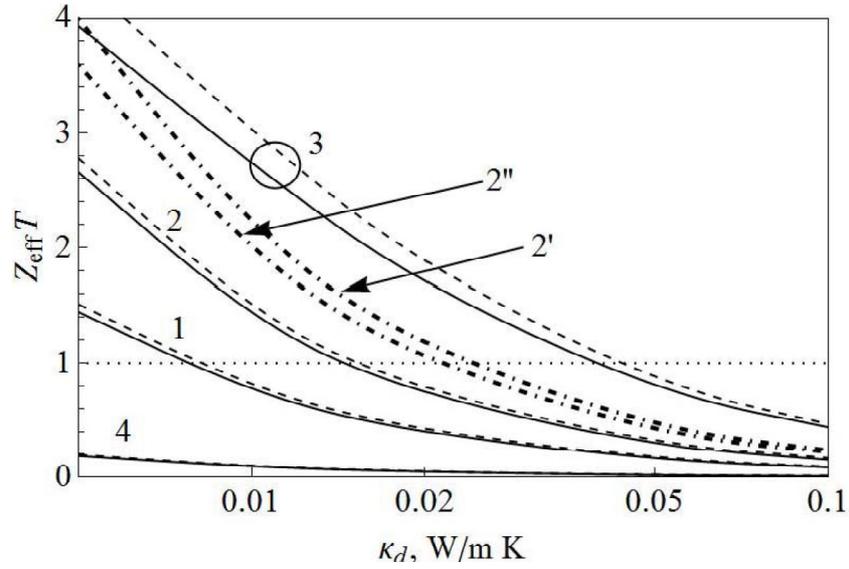

Fig. 5. The dependence of the thermoelectric figure of merit on dielectric thermal conductivity for different material parameters. Solid and dashed curves are plotted for semiconducting and metallic nanoparticles correspondingly assuming the same effective mass in nanoparticle and barrier. Dash-dotted curves are plotted taking into account the influence of anisotropy of semiconducting material (see text for details). Barrier parameter $d$ and $\varepsilon_b$ are the following: 1 - 2nm, 0.1eV; 2, 2', 2'' – 5nm, 0.1eV; 3 – 5nm, 0.05eV; 4 – 5nm, 0.2eV.

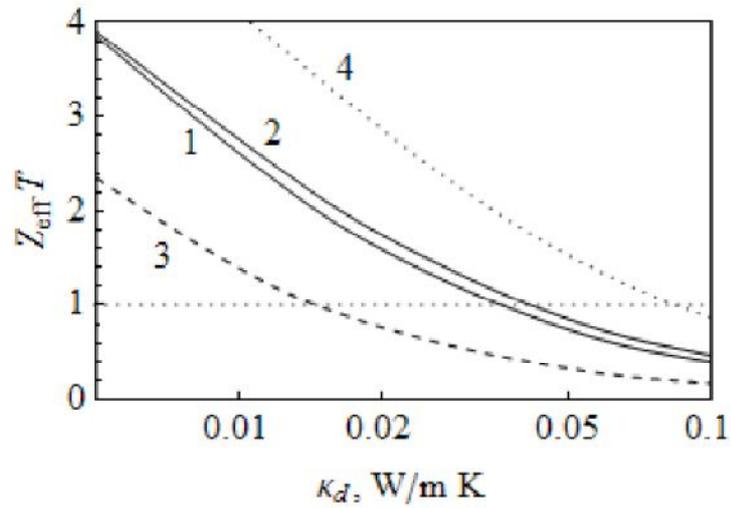

Fig. 6. The dependence of the thermoelectric figure of merit on dielectric thermal conductivity for semiconducting grains and different geometric parameters. Barrier parameters $\varepsilon_b = 0.05\,\text{eV}$, $d = 5\text{nm}$. 1 and 2 - $\theta = 15^\circ$, $2a = 30\text{nm}$ and $10\text{nm}$; 3 - $\theta = 30^\circ$, $2a = 10\text{nm}$; 4 – layered geometry $2a = 10\text{nm}$.